\documentclass[5pt,final]{iopart}
\expandafter\let\csname equation*\endcsname\relax
\expandafter\let\csname endequation*\endcsname\relax
\usepackage{amsmath}

\usepackage{amssymb}

\usepackage[makeroom]{cancel}

\usepackage{mhchem}

\usepackage{graphicx}

\usepackage{dcolumn}

\usepackage{bm}
\usepackage[normalem]{ulem}
\usepackage{color}

\usepackage{verbatim}

\usepackage{epstopdf}

\newcommand{\bff}{\mathbf{f}}

\newcommand{\bbr}{\mathbf{r}}

\newcommand{\tri}{\triangle}

\newcommand{\sep}{ \ \ \ , \ \ \ }

\newcommand{\beq}{\begin{equation}}
\newcommand{\eeq}{\end{equation}}
\newcommand{\beqn}{\begin{eqnarray}}
\newcommand{\eeqn}{\end{eqnarray}}

\newcommand{\dd}{{\rm d}}
\newcommand{\ee}{{\rm e}}
\newcommand{\eq}{Eq.\ }

\newcommand{\eqs}{Eqs }
\newcommand{\fig}{Fig.\ }

\newcommand{\la}{\langle}

\newcommand{\ra}{\rangle}

\begin{document}

\title[ Equilibrium kinetics of polymerisation]{Equilibrium kinetics of self-assembling, semi-flexible polymers 
}

\author{Chiu Fan Lee}
\address{Department of Bioengineering, Imperial College London, South Kensington Campus, London SW7 2AZ, U.K.}
\ead{c.lee@imperial.ac.uk}
\begin{abstract}
Self-assembling, semi-flexible polymers are ubiquitous in biology and technology. However, there remain conflicting accounts of the   equilibrium kinetics for such an important system. Here, by focusing on a dynamical description of a minimal model in an overdamped environment, I identify the correct kinetic scheme that describes the system at equilibrium in the limits of high bonding energy and dilute concentration.
\end{abstract}
\submitto{\JPCM}
\maketitle

\section{Introduction}
Many polymerisation processes are in principle reversible -- thermal fluctuations will inevitably break a polymer apart and two  polymers can potentially join up upon encountering each other.  Indeed, there are many examples of synthetic and natural polymers that remodel themselves by breakage and associations at an experimentally accessible scale \cite{cates_jpcm90,schoot_r05,bolisetty_biomacromol12}. Held at fixed temperature, these re-modelling systems will eventually reach thermal equilibrium. However, it is surprising that we still lack a dynamical picture that describes reversible polymerization at equilibrium based on first principles. 
In the literature, the kinetic schemes of polymerisation are typically postulated in  ad hoc manners, with contradictory assumptions \cite{hill_biophysj83, krapivsky_b10}. 
Here, I will identify the physically correct kinetic scheme for  a minimal model of self-assembling, semi-flexible polymers in the limits of high bonding energy and dilute polymer concentration.

\section{A minimal model}
\begin{figure}
	\begin{center}
		\includegraphics[scale=.38]{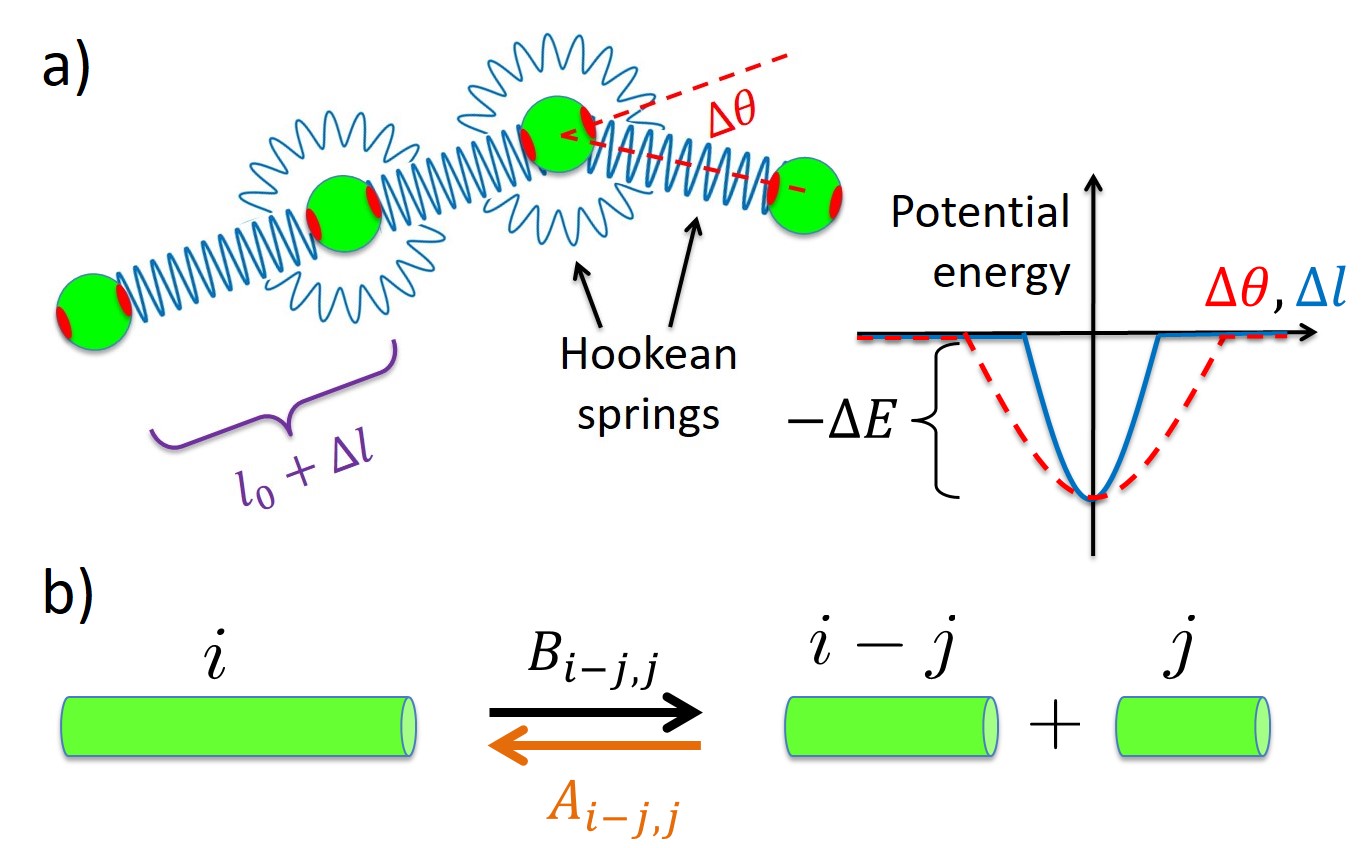}
	\end{center}
	\caption{a) {\it A minimal model of self-assembling semi-flexible polymers.} Monomers (green beads) bind to each other via the red patches at the polar ends. The length and rigidity of the resulting polymer are regulated by the potential energy functions that are quadratic for $|\tri \theta| \leq \Theta_c$ and $|\tri l| \leq L_c$. b) {\it Schematics of the association and breakage events.} Due to thermal perturbations, $i$-mers in the system  will be broken up into pairs of $(i-j)$-mers and $j$-mers with net rate $B_{ij}c_i$ per unit volume where $c_i$ is the $i$-mer concentration; the reverse can also happen with net rate $A_{i-j,j} c_{i-j}c_j$ per unit volume.
	}
	\label{fig:pic1}
\end{figure}

At the simplest level, a self-assembling polymer system can be viewed as a collection of particles that can bind to each other due to some short-ranged potential energy function. To represent such a picture in a minimal way, I consider a 
collection of particles (monomers), each having two sticky patches at two polar ends (\fig \ref{fig:pic1}a). The sticky patches are assumed to be small so that branching is not possible. Namely, all polymers are linear. 
 The interactions of the particles are described by two quadratic energy functions constraining the bond  length and  polymer rigidity,
with the cut-off on the stretching given by $L_c$ and that of bending by $\Theta_c$, respectively (\fig \ref{fig:pic1}a). Specifically, the overall potential energy for a $n$-mer (a polymer consisting of $n$ monomers) is
\beq
\label{eq:energy}
U_n(\{ \bbr\})= \tri E \left\{
\sum_{k=1}^{n-1}\frac{\triangle l_k^2}{L_c^2} +
\sum_{h=2}^{n-1} \frac{\triangle \theta_h^2}{\Theta_c^2} -(n-1)
\right\}
\ ,
\eeq
where the distance between the $k$-th bead and the $(k+1)$-bead is denoted by $l_0+\triangle l_k$, and the bending angle $\tri \theta_h$ at the $h$-th bead is
\beq
\tri \theta_h = \arccos \frac{ (\bbr_{h+1}-\bbr_h) \cdot (\bbr_{h}-\bbr_{h-1})}{ |\bbr_{h+1}-\bbr_h ||\bbr_{h}-\bbr_{h-1}|}
\ ,
\eeq
with $\bbr_h$ being the position of the $h$-th particle in the polymer. 
The first term in (\ref{eq:energy}) controls the 
the distances between the connected beads and the second term enforces the rigidity of the polymers, with $\Theta_c$ small enough that loop formation in the system can be ignored 
(\fig \ref{fig:pic1}a).  Note that the  stretching part and the bending part of the potential energy are non-zero only if $|\tri l_k| <L_c$ for $k=1, \dots, n-1$, and  $|\tri \theta_h| < \Theta_c$ for  $h=2,\ldots,n-1$. 
The last term in  (\ref{eq:energy})  
leads to a lowering of the system's energy by $(n-1) \tri E$ when a $n$-mer forms. Increasing $\tri E$ thus promotes polymerisation. 
Note that additional volume exclusion interactions can be added to the system, however, the specificity of such interactions are unimportant to our discussion here.

One advantage of this minimal model is that the free energy of the system can be calculated in the dilute limit  via a mean-field method (Appendix A). The key results are that at thermal equilibrium, the concentration of $i$-mers in the system is:
\beq
\label{eq:ci}
c_i = P \exp \left(-\frac{i}{\bar{n}}\right) \sep {\rm for \ } i>1\ ,
\eeq
where  $\bar{n}$ is the average size of the polymers in the system,
while the monomeric concentration $c_1$ approaches the fixed threshold concentration $C_c$  asymptotically as the total particle concentration $C_{\rm tot}$ increases.  The expression of the threshold concentration is (\ref{eq:ccB}):
\beq
\label{eq:cc}
C_c = \frac{4(\beta \tri E)^{5/2}}{\pi^{3/2} l_0^2 L_c \Theta_c^4} \ee^{-\beta \tri E} 
\ . 
\eeq
These results are of course consistent with well known analytical results in self-assembling polymeric systems \cite{cates_jpcm90,
carl_jphys97,israelachvili_b10}, and have also been confirmed by direct simulation methods \cite{kroger_pre96,carl_jphys97, sciortino_jpcm08}.

In the case of $C_{\rm tot} \gg C_c$, most particles are in the polymeric form. 
In this limit (\ref{eq:barn3}):	
\beq
\label{eq:barn}
\bar{n} 
=\frac{ \pi^{3/4} l_0 L_c^{1/2} \Theta_c^{2}}{2  (\beta \tri E)^{5/4}} \sqrt{C_{\rm tot} \ee^{\beta\tri E}}
\ .
\eeq
This is again consistent with the well-known result that the average aggregate size $\bar{n}$ scales like $\sqrt{ C_{\rm tot} \ee^{\tri E/k_BT}}$ \cite{cates_jpcm90,
	carl_jphys97,israelachvili_b10}.
Note that the degree of polymerization is  controlled by both the total concentration of particles in the system and the bonding energy.

Because of  \eq (\ref{eq:barn}) and the overall particle number conservation, the prefactor in (\ref{eq:ci}) has the following form: 
\beq
\label{eq:P}
P 
 =
\frac{C_{\rm tot}}{\bar{n}^2} = \frac{4 (\beta \tri E)^{5/2}}{ \pi^{3/2} l_0^2 L_c \Theta_c^{4}}   \ee^{-\beta \tri E}
\ .
\eeq
Given (\ref{eq:ci}) \& (\ref{eq:P}), the polymer concentration $C_{\rm poly}$ in the system can be calculated to be:
\begin{subequations}
\beqn
C_{\rm poly} &=& 
P \int_0^\infty \dd i \exp(-i/\bar{n}) =\bar{n} P 
\\
\label{eq:cpoly}
&=&\frac{2  (\beta \tri E)^{5/4}}{ \pi^{3/4} l_0 L_c^{1/2} \Theta_c^{2}} \sqrt{C_{\rm tot} \ee^{-\beta\tri E}}
\ .
\eeqn 
\end{subequations}

Here, besides concentrating on the regime of high level of polymerisation ($C_{\rm tot} \gg C_c$), I will focus on the dilute regime in which the polymers are far apart \cite{doi_b86}.  Since the typical length of a polymer is $\bar{n} l_0$, the dilute regime corresponds to $\bar{n}^3l_0^3 C_{\rm poly} \ll 1$. From  (\ref{eq:barn}) and (\ref{eq:cpoly}), the restriction can be re-written as $l_0^3 C_{\rm tot} \ll \ee^{-\beta \tri E/2}$.  Since $l_0^3C_c \sim \ee^{-\beta \tri E}$  (\ref{eq:cc}), the two restrictions of having a high degree of polymerisation {\it and} being in the dilute limit can be satisfied if 
\beq
\ee^{-\beta \tri E} \ll l_0^3C_{\rm tot} \ll \ee^{-\beta \tri E/2}\ .
\eeq

\begin{figure}
	\begin{center}
		\includegraphics[scale=.6]{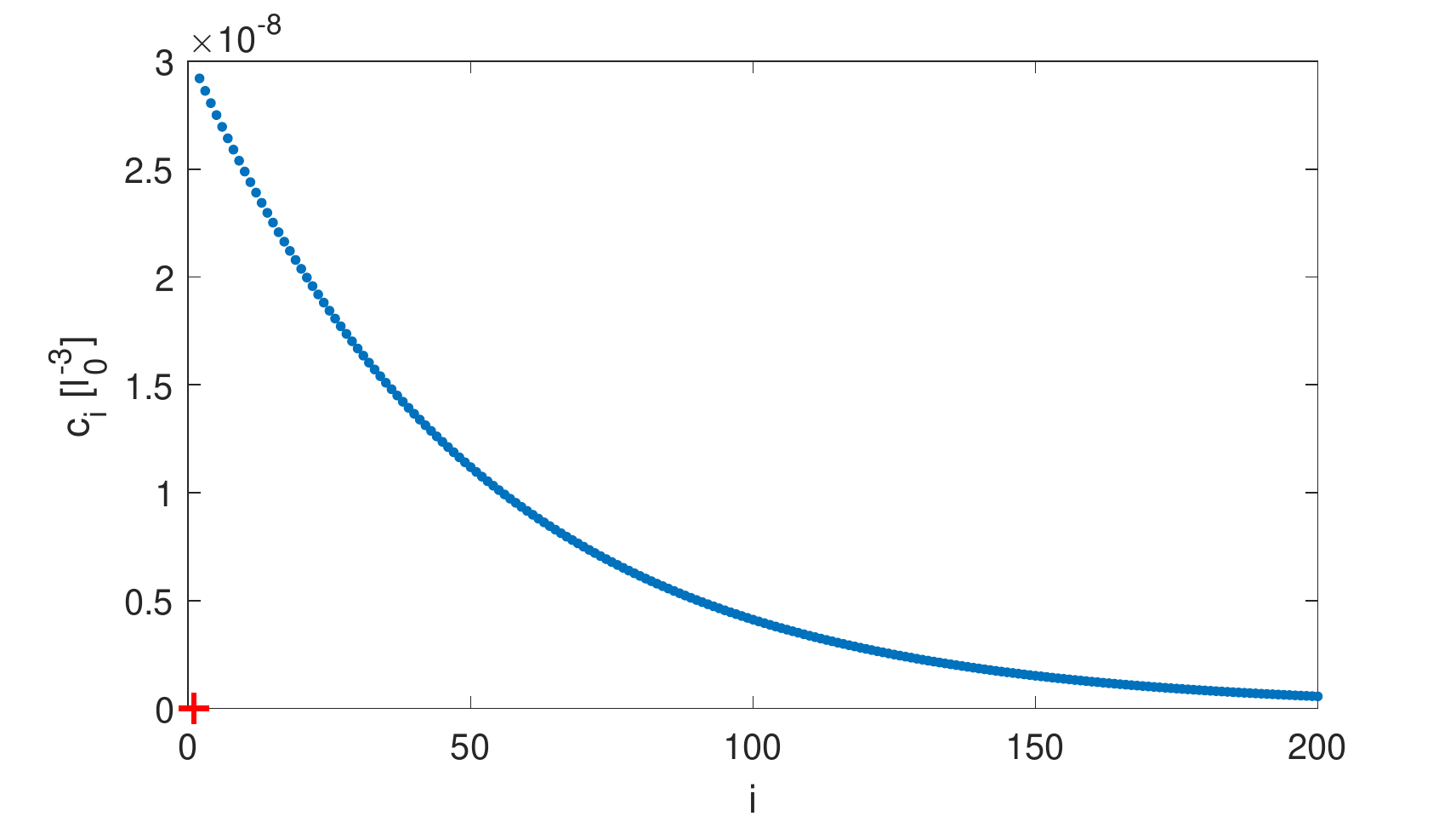}
		
	\end{center}
	\caption{{\it Length distribution of self-assembling polymers at equilibrium.} The distribution at equilibrium is an exponential function of $i$, the size of the polymer. The parameters are $L_c = l_0/10$, $\Theta_c = 0.1$ rad, $\tri E = 40k_BT$,  $C_{\rm tot} = 7.6\times 10^{-5} l_0^{-3}$. As a result, the average size $\bar{n}$ is about  50.		
		The monomer concentration, $c_1 \approx C_c$, is denoted by the red cross. 
	}
	\label{fig:dist}
\end{figure}

\subsection{An example}
\label{sec:ex}
Letting $L_c = l_0/10$, $\Theta_c = 0.1$ rad $\approx 5.7^\circ$, and $\tri E = 40k_BT$, one finds that the threshold concentration $C_c$ is $6.1 \times 10^{-13} l_0^{-3}$. If the total solute concentration of $C_{\rm tot} = 7.6\times 10^{-5} l_0^{-3} \gg C_c$, which corresponds to the solute's volume fraction of around $4.0\times 10^{-5}$ (assuming each particle has a volume of $\pi l_0^3/6$), the average size $\bar{n}$ is then 50  and the polymer concentration $C_{\rm poly}$ is $1.5 \times 10^{-6} l_0^{-3}$. Since the typical length of the polymer is $\bar{n} l_0 =50l_0$, each polymer in the isotropic phase will occupy a typical volume of $4\pi (\bar{n} l_0/2)^3/3 \approx 6.5\times 10^4 l_0^3$. The polymeric system is therefore in the dilute phase since $5.2\times 10^5 l_0^3 \times C_{\rm poly}\approx 0.098$. In this system, the monomer concentration $c_1 \approx C_c$ is negligible (\fig \ref{fig:dist}). Indeed, $c_1$ only surpasses $c_i$ for $i>540$.

\section{Equilibrium kinetics}
I have so far described the static configuration of the system. To consider the system's kinetics, I will further assume that the particles are in an over-damped environment so that each particle exhibits Brownian motion. For instance, for the $k$-th particle in an $n$-mer, the equation of motion is
\beq
\frac{\dd \bbr_k}{\dd t} = -\frac{1}{\zeta} {
	\nabla}_{\bbr_k}  U_n + \sqrt{ \frac{2k_BT}{\zeta}} \bff_k
\eeq
where the potential function $U_n$ is described by (\ref{eq:energy}), $\zeta$ is the damping coefficient and $\bff_k$ denotes three dimensional independent Gaussian noises with zero mean and unit variance. 

Driven by thermal perturbations, the monomers' positions fluctuate and a polymer will eventually be broken either because a particular angle goes beyond the threshold angle $\Theta_c$, or because a particular length gets beyond the threshold length $L_c$. On the other hand, two polymers can rejoin if one polymer's end encounter the other's at the right distance and angle. As a result, even though the exponential size distribution does not change with time, the breakage and association events happen continuously and the whole  system is in  a dynamical equilibrium.

In the dilute limit, the kinetic equations that describe the evolution of the set of concentrations $\{ c_i\}$ can be written generically as \cite{krapivsky_b10}
\beq
\label{eq:ss}
\frac{\dd c_k}{\dd t}=
\frac{1}{2}\sum_{i+j=k} {A_{ij}} c_i c_j -c_k \sum_{j \geq 1} {  A_{kj}} c_j
+\sum_{j \geq 1} {B_{kj} } c_{j+k} -\frac{c_k}{2} \sum_{i+j=k} { B_{ij}}
\ ,
\eeq
where $A_{ij}c_ic_j$ describe the association rates of pairs of $i$-mers and $j$-mers per unit volume, while $B_{ij}$ corresponds to the breakage rate of a $(i+j)$-mer into a $i$-mer and an $j$-mer (\fig 
\ref{fig:pic1}b). Note that the above set of kinetic equations is completely general and there are only two underlying assumptions: 1) third-body interactions can be ignored due to the dilute-limit condition ($C_{\rm poly}l_0^3 \ll 1$) , and 2) the breakage of a polymer is independent of the other polymers around it, which is again motivated by the dilute-limit condition.

To describe the kinetics of polymerization from first principles, one thus need to calculate the sets of $\{ A_{ij} \}$ and $\{B_{ij}\}$ for $i,j \in {\bf N}$. This task is drastically simplified at thermal equilibrium since the detailed balance conditions dictate that \cite{krapivsky_b10}:
\beqn
A_{ij} c_i c_j &=& B_{ij} c_{i+j}
\\
\label{eq:DB}
PA_{ij}&=&B_{ij}
\eeqn
where the second equality comes from using the equilibrium distribution of $c_i$ (\ref{eq:ci}). In other words, specifying 
$\{ A_{ij} \}$ {\it or} $\{B_{ij}\}$ suffices to determine completely the kinetic scheme at thermal equilibrium.

\subsection{Two scenarios: Smoluchowski scheme vs.\ uniform breakage scheme} 
There are two prevailing kinetic schemes of polymerization at equilibrium. The first one assumes that $A_{ij} =A$ and $B_{ij}=B$, i.e., both association and breakage events are independent of the sizes of the polymers involved \cite{cates_jpcm90,krapivsky_b10}.  The {\it uniform breakage scheme} has been typically postulated in an {ad hoc} manner, motivated  mainly by its analytical tractability.

In the second kinetic scenario \cite{hill_biophysj83}, the rates of polymer association are calculated by assuming that association events are diffusion-limited (the {\it Smoluchowski scheme}). Using this approach, it is natural to conclude that $A_{ij}$ would be smaller than $A_{kh}$ if both  $i$ and $j$ are greater than $h$ and $k$  since  long polymers diffuse less quickly than shorter ones. In other words, $\{ A_{ij} \}$ and thus also $\{ B_{ij} \}$ depend on indices $i$ and $j$, which is of course incompatible with the {\it uniform breakage scheme}.

Recently, I have proved mathematically that in the asymptotic limit of high bonding energy, the breakage propensity of each bond in a freely diffusing polymer is independent of the location and the length of the polymer, irrespectively of whether breakage is by extension \cite{lee_thermal09} or by bending  \cite{lee_jpcm15}.  These results demonstrate that for a single polymer, $B_{ij}$  is indeed independent of the indices $i,j$. If the  kinetics of polymerisation is correctly described by \eq (\ref{eq:ss}), then the above result clearly supports the {\it uniform breakage scheme}, 
 at least in the asymptotic limit of high bonding energy. In the following, I will discuss why the Smoluchowski scheme is incorrect.

\vspace{.1in}
\subsection{What's wrong with the Smoluchowski scheme} 
In the Smoluchowski picture, the association of two polymers, of sizes $i$ and $j$, say, are assumed diffusion-limited and the rate can for instance be calculated as follows: the $i$-mer is held stationary at the origin and the concentration profile of the $j$-mer results from solving the diffusive equation such that the concentration is zero if the configurations allow the  two polymers to join up; while  the concentration is non-varying and equal to $c_j$  far from the origin \cite{hill_biophysj83,jackson_b06}. 

I will now show that such an approximation is problematic because it turns out that in the high $\tri E$ limit, {\it almost all} polymers will break and re-join with their fragments many times over before diffusion takes them to other polymers far away.
To demonstrate this, I will first consider the association kernel for polymers $i$ and $j$ ($A^{\rm (Smol)}_{ij}$), and then obtain the corresponding breakage rate ($B^{\rm (Smol)}_{ij}$) via the detailed balance condition (\ref{eq:DB}). This will be compared with the correct asymptotic ($\tri E \rightarrow \infty$)  breakage rate $B=B_{ij}$ calculated previously  \cite{lee_thermal09,lee_jpcm15}.

Starting with the Smoluchowski picture, the association rate per unit volume has been calculated by Hill \cite{hill_biophysj83}: 
\beq
A^{\rm (Smol)}_{ij} c_ic_j 
= \frac{\pi k_BT L_c^2 \Theta_c^2}{4 l_0 \zeta}  \frac{j\log i+i \log j}{ij(i+j)} c_ic_j
\ ,
\eeq
which  gives, via the detailed balance condition in  (\ref{eq:DB}), the breakage rate as follows:
\beq
B^{\rm (Smol)}_{ij} =A^{\rm (Smol)}_{ij}P = \frac{\pi^{3/2} L_c}{ l_0^3 \Theta_c^{2}} \frac{k_BT}{\zeta}  \frac{j\log i+i \log j}{ij(i+j)}  (\beta \tri E)^{5/2}\ee^{-\beta \tri E}  
\ .
\eeq

On the other hand, the breakage rate by bending due to thermal perturbations  in the limit $\tri E \rightarrow \infty$ is \cite{lee_jpcm15}: 
\beq
B^{\rm (bend)} =
\frac{24}{ l_0^2\Theta_c^2 } \frac{k_BT}{\zeta}  (\beta \tri E)^{2}  \ee^{-\beta\tri E}
\eeq
and the corresponding breakage rate by extension is \cite{lee_thermal09}:
\beq
B^{\rm (ext)}= \frac{8}{\sqrt{\pi} L_c^2 } \frac{k_BT}{\zeta}  (\beta \tri E)^{3/2} \ee^{-\beta\tri E}
\ .
\eeq


Let us now consider the following ratios: 
\begin{subequations}
	\label{eq:ratios}
\beqn
\frac{B^{\rm (Smol)}_{ij} }{B^{\rm (bend)} } &=& \frac{\pi^{3/2} L_c}{24 l_0 }   \frac{i\log j+j \log i}{ij(i+j) } \sqrt{ \beta \tri E}
\\
\frac{B^{\rm (Smol)}_{ij} }{B^{\rm (ext)} } &=& \frac{\pi L_c^3}{8 l_0^3 \Theta_c^2} \frac{i\log j+j \log i}{ij(i+j) } \beta \tri E 
\ .
\eeqn
\end{subequations}
Superficially, $B^{\rm (Smol)}$ dominates over $B^{\rm (bend,\ ext)}$ when $\tri E$ becomes large, however since the typical polymer length also grows exponentially with $\tri E$, the ratios in \eqs (\ref{eq:ratios}) for {\it almost all} pairs of polymers go to zero exponentially rapidly as $\tri E$ grows. 
For instance, for two polymers of the average size $\bar{n}$:
\begin{subequations}
	\label{eq:ratio}
\beqn
\frac{B^{\rm (Smol)}_{\bar{n} \bar{n}} }{B^{\rm (bend)} } &=& \frac{\pi^{3/2} L_c \sqrt{\beta\tri E}  }{24 l_0 } \frac{ \log  \bar{n}}{ \bar{n}^2} 
\\
\label{eq:ratio3}
&\sim & \left(\beta\tri E \right)^2  \ee^{-\beta\tri E} \rightarrow_{\tri E \rightarrow \infty} 0
\ ,
\eeqn
\end{subequations}
where (\ref{eq:ratio3}) follows from (\ref{eq:barn}). 

More generally, the expressions in (\ref{eq:ratios}) indicates that both ratios are negligible if $\min(i,j) \gg \beta \tri E$. Since given any integer $k$, the fraction of polymers of sizes smaller than $k$ is $C_{\rm tot}^{-1} \int_0^k c_i \dd i = 1-\ee^{-k/\bar{n}} \approx k/\bar{n} \sim k \ee^{-\beta \tri E/2}$, we see that as $\beta \tri E \rightarrow \infty$, most polymers in the system are of sizes of order $\ee^{-\beta \tri E/2}$. In other words, the ratios in  (\ref{eq:ratios}) go to zero exponentially quickly for almost all pairs of polymers in the system. Rephrasing this more mathematically, for any small and positive number $\epsilon$, there exists a threshold $\beta \tri E$ beyond which the ratios in (\ref{eq:ratios}) are smaller than $\epsilon$ for $(1-\epsilon)$ fraction of  all polymer pairs in the systems. This result  shows that  the breakage rate as obtained in the Smoluhowski picture is typically negligible in the high bonding energy limit compared to the actual breakage rate. Importantly, this does not merely suggests that the contribution from the Smoluchowski scenario can be ignored, instead, it points to the fact that in an association event, the far field limit being the dominant source is incorrect. Rather, the dominant source originates from the very ends of the polymers concerned. I will illustrate this dynamical picture in a simplified
model in section \ref{twopolymermodel}.

To summarise this section, by comparing the relative magnitude of the breakage rates calculated from the Smoluchowski and the uniform breakage schemes, I have demonstrated that  the equilibrium kinetics of the minimal system considered here is correctly described by the uniform breakage scheme, in the asymptotic limit of $\beta \tri E \rightarrow \infty$.

\begin{figure}
	\begin{center}
		\includegraphics[scale=.6]{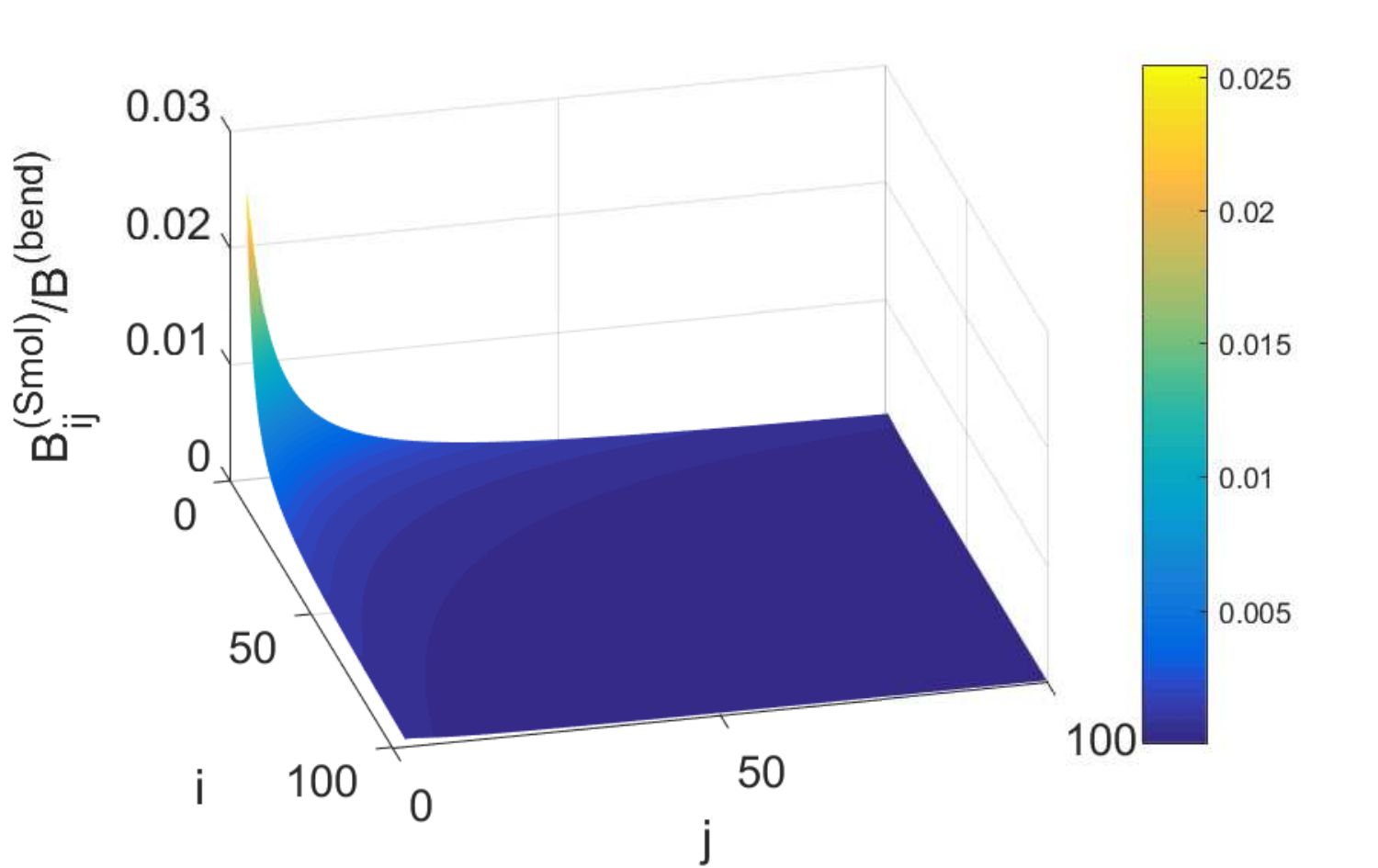}
		
	\end{center}
	\caption{{\it The dominance of the uniform breakage scheme.} The ratio of the breakage rate from the Smoluchowski scheme and that from the uniform breakage scheme (\ref{eq:ratios}a) as a function of the resulting polymers of sizes $i$ and $j$.
	}
	\label{fig:contour}
\end{figure}

\subsection{Equilibrium kinetics of example \ref{sec:ex}}
As an example, I will now consider the ratios in (\ref{eq:ratios}) with the concrete example introduced in section \ref{sec:ex}.
Using again the parameters in Sect.~\ref{sec:ex}, $B^{\rm (bend)}/B^{\rm (ext)} \approx 34$ and so the breakage of a free polymer is predominantly via bending. In particular, if we assume the system is a collection of colloid of diameters $l_0=10$nm in water, the damping coefficient  $\zeta$ is, via the Einstein-Stokes relation, $6\pi\eta l_0$ where $\eta$ is the dynamic viscosity of water, the resulting breakage rate is then $B^{\rm (bend)} \approx 8.0\times 10^{-6}$ per second at $T=300$K. The expression (\ref{eq:ratios}a) is shown in \fig \ref{fig:contour}, demonstrating that the ratio is in fact always smaller than one for all pairs of $i$ and $j$ even when $\beta \tri E = 40$. Indeed, given that the cut-off $L_c$ is expected to be smaller than the actual particle separation $l_0$, $\beta \tri E$  generically has to be very large in order for the ratios in \eq (\ref{eq:ratio}) to be greater than 1. For instance, in the present case where $L_c/ l_0 =0.1$ and $\Theta_c=0.1$, even for $i=j=2$, the ratios 
are greater than 1 only if $\beta \tri E > 61890$ in (\ref{eq:ratio}a) and $\beta \tri E > 147$ in (\ref{eq:ratio}b).

%

\section{A simplified three-species model}
\label{twopolymermodel}
To clarify the dynamics of breakage and association, I will now consider a simplified model where there are only three types of polymers in the system: dimers, $n$-mers where $n \gg 1$, and $(n+2)$-mers. In this simplified three-species model, one can again minimise the free energy of the system, similar to what is done in Appendix A, in order to calculate the concentrations of three distinct types of polymers: $c_2$, $c_n$ and $c_{n+2}$. I will now use one of the end beads of the $n$-mer as our reference frame, i.e., our reference frame diffuses with the end bead. We have seen that as far as the association event is concerned, the source of the dimer is not from the far field, but rather from the breakage of the $(n+2)$-mer that created the two polymers in the first place. As a result, the source of the dimer is in fact at the ends of the $n$-mer, and upon breaking away, it diffuses around until it gets reabsorbed by the $n$-mer to form a $(n+2)$-mer, or potentially less likely, diffuses to the surrounding of another $n$-mer far away and gets reabsorb there. The actual concentration field of the dimer will of course be highly dependent on the specific model used, such as how the steric  interactions are modelled, however the underlying physics  is universal: the concentration field at equilibrium of a dimer around an $n$-mer corresponds to solving the diffusion equation subject to the following boundary conditions: 1) the source comes from the breakage point at the two ends of the $n$-mer centred in a box of volume $1/c_n$ with no-flux boundary condition at the boundary; 2) there is a absorbing boundary that corresponds to the event when the dimer and the $n$-mer can re-join to form an $(n+2)$-mer; and
3) any further boundary conditions due to the steric interactions between the two polymers. 

An example of the equilibrium spatial distribution of a dimer around an $n$-mer is illustrated in \fig \ref{fig:dist}. Here, the system is two-dimensional (2D) and the reference frame is chosen such that the long $n$-mer ($n \gg 1$) is held fixed vertically. As aforementioned, the dominant source of the dimer is at the end of the $n$-mer, i.e., the dimer is produced through breaking off from the $n$-mer. It then diffuses around within the bounding box until it re-attaches with the $n$-mer to form a $(n+2)$-mer. The average orientation of the dimer around the $n$-mer is also shown in \fig \ref{fig:dist}b.

\begin{figure}
	\begin{center}
		\includegraphics[scale=.85]{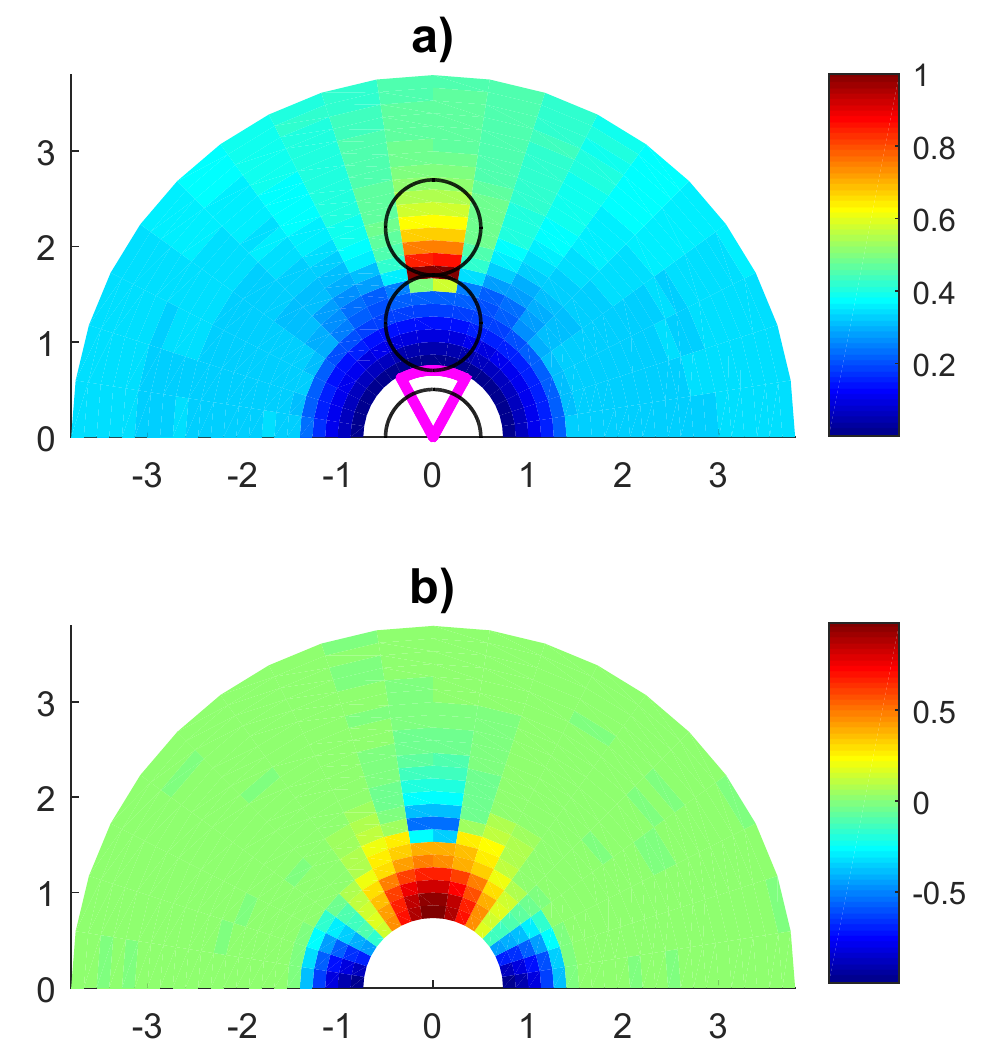}
	\end{center}
	\caption{{\it The density distribution (a) and the average orientation (b) of a dimer around the end of another polymer in the simplified model described in section \ref{twopolymermodel}.} The dimer (black circles) is ejected through extensile breakage from the end of a polymer positioned vertically below. Upon ejection the dimer diffuses until rejoining the end of the polymer when the distance of the centre of one of beads is of distance $l_0+L_c$ from the origin and when the angle between the length of the polymer and the vertical axis is between $\Theta_c$ (region indicated by the the purple boundary). Hard-core repulsion between the beads exist with a distance of $l_0$, hence the forbidden area (white) around the origin.
		a) The normalized density map of the centre of mass of the dimer.  b)		
		The average orientation $\la (2\cos \phi^2 -1) \ra$ of the dimer, where $\phi$ is the angle between the length of the dimer and the horizontal axis. Thus, the average orientation is 0 when the orientation is isotropic, 1 when it is strictly horizontal and $-1$ when it is vertical.  Simulation details are provided in Appendix B.
	}
	\label{fig:dist}
\end{figure}

\section{Summary \& Outlook}
By focusing on an analytically tractable model of a self-assembling polymeric system, I have demonstrated that the equilibrium kinetics of the system in the limits of  high bonding energy and dilute concentration is well described by the uniform breakage scheme. 
 Although the model considered assumes a specific form of the potential energy functions for simplicity, the same conclusion is expected to hold for more general energy functions since the breakage propensity remain uniform even when the binding energy functions are not quadratic \cite{lee_thermal09,lee_jpcm15}.  

 Shifting down from the high bonding energy regime, the picture is unclear. First, the breakage propensity may no longer be uniform across a polymer \cite{lee_thermal09,lee_jpcm15}; second, the ratios in (\ref{eq:ratios}) may no longer be negligible for the majority of the polymer pairs in the system. Therefore, what the quantitative model in this regime is regarding the equilibrium kinetics remains an interesting open problem.

\appendix
\section{Equilibrium configuration}
In this appendix, I will review how the equilibrium configuration can be obtained from minimising the free energy of the system,  which is a summary of the analytical calculations presented in \cite{liu_r17}. Note that a similar calculation has also been performed for self-assembling, flexible polymers in \cite{carl_jphys97}.

To calculate the polymer length distribution at thermal equilibrium, I will start with the total partition function 
\begin{eqnarray}
\label{eq:Ztot}
Z_{\rm tot} =\prod_{i}'\frac{1}{N_i!} \left(\frac{Vz_i}{\Lambda^3}\right)^{N_i},
\end{eqnarray}
where $N_i$ is the number of $i$-mers in the system with volume $V$,
and $\Lambda$ is an immaterial constant of dimension $V^{1/3}$ to render the partition function dimensionless.  The prime above the product sign refers to the particle number conservation condition: $\sum_i i N_i = N_{\rm tot}$ where $N_{\rm tot}$ is the total number of particles in the system. Furthermore, $z_i$ is the 
configurational partition function of an $i$-mer such that the position of the first particle in the polymer is fixed, which I will calculate later.

To study the equilibrium configuration of the system, I minimise the total free energy density:
\begin{eqnarray}
f_{\rm tot} = \frac{F_{\rm tot}}{V} = -\frac{1}{\beta V}\ln Z_{\rm tot}
\ ,
\end{eqnarray}
with respect to the polymer number distribution $\{N_i\}$ conditional on the overall conservation of particle number. 

Focusing on a polymeric system with a large average size, one can approximate $f_{\rm tot}$ as a functional of $\{ n_i \}$:
\beq
f_{\rm tot}[\{n_i\}] \approx  \beta^{-1}\int_0^\infty  n_i \left[\ln (\Lambda^3 n_i)  -\chi i +\xi-1\right] \dd i
\ ,
\eeq
where $n_i \equiv N_i/V$ and 
\beq
\label{eq:zi}
\ln z_i = \chi i -\xi
\ . 
\eeq

By minimizing the above functional subject to the number conservation condition $\int_0^\infty i n_i \dd i = N_{\rm tot}/V \equiv C_{\rm tot}$, one finds
\beq
\label{eq:A5}
n_i  = \frac{\ee^{ -i/\bar{n}}}{\Lambda^3 \ee^\xi}
\eeq
where
\beq
\label{eq:barn2}
\bar{n} = \sqrt{\Lambda^3  \ee^{\xi} C_{\rm tot}}
\ .
\eeq
To obtain an analytical expression for $\xi$, we need to calculate the $i$-mer partition function $z_i$, which is of the form:
\beqn
\label{A1}
z_i&=&\frac{4\pi l_0^2  }{\Lambda^{3(i-1)}} {{\rm e}}^{(i-1)\beta \tri E}\left[\int_{-l_c}^{l_c} \exp \left(-\frac{\beta \tri E \triangle l^2}{L_c^2} \right) \dd \triangle l  \right]^{i-1}
\\
&&
\label{A2}
\times
\left[2\pi l_0^2\int_{0}^{\theta_c}  \sin (\triangle \theta )\exp \left(-\frac{\beta \tri E \triangle \theta^2}{\Theta_c^2} \right) \dd \triangle \theta\right]^{i-2}
\ ,
\eeqn
where the term $4\pi l_0^2$ comes from the surface area swept out by the second bead connected to the first bead which is assumed fixed in position, and the exponential term results from the $(i-1)$ bonds in the $i$-mer. The term in the first squared brackets comes from the extensile degree of freedom in each bond, and the term in the second squared brackets comes from the bending degree of freedom between every pair of consecutive bonds.

Since $\beta \tri E$ is assumed to be large, I will approximate $z_i$ by extending the limits of the integrals as follows:
\begin{subequations}
\begin{eqnarray}
z_i
&\approx &\frac{4\pi l_0^2 {{\rm e}}^{(i-1)\beta \tri E} }{\Lambda^{3(i-1)}}\left[\int_{-\infty}^{\infty} \exp \left(-\frac{\beta \tri E \triangle l^2}{L_c^2} \right) \dd \triangle l  \right]^{i-1}
\\
&&
\times
\left[2\pi l_0^2\int_{0}^{\pi}  \sin (\triangle \theta )\exp \left(-\frac{\beta \tri E \triangle \theta^2}{\Theta_c^2} \right) \dd \triangle \theta\right]^{i-2}\nonumber\\
\label{Aans}
&=&\frac{4\pi l_0^2 {{\rm e}}^{(i-1)\beta \tri E} }{\Lambda^{3(i-1)}}\left(\frac{\pi L_c^2}{\beta \tri E}  \right)^{(i-1)/2}
\left( \frac{\pi l_0^2 \Theta_c^2}{\beta \tri E}
\right)^{i-2}
\ ,
\end{eqnarray}
\end{subequations}
where in (\ref{Aans}), only the leading order term in $(\beta \tri E)^{-1}$ is kept for the integral w.r.t.~$\tri \theta$. 

Since $\ln z_i = \chi i-\xi$ by definition (\ref{eq:zi}), one can see that 
\beq
\xi =\beta \tri E-\ln \frac{4 \Lambda^3 (\beta \tri E)^{5/2}}{\pi^{3/2} L_c l_0^2 \Theta_c^4}
\ .
\eeq
Given this expression, we can finally find from (\ref{eq:barn2}) that
\beq
\label{eq:barn3}
\bar{n} =\frac{ \pi^{3/4} l_0 L_c^{1/2} \Theta_c^{2}}{2  (\beta \tri E)^{5/4}} \sqrt{C_{\rm tot} \ee^{\beta\tri E}}
\ .
\eeq
The monomer threshold concentration $C_c$ corresponds to the concentration $C_{\rm tot}$ at which $\bar{n}=1$ \cite{liu_r17}. Therefore,  
\beq
\label{eq:ccB}
C_c = \frac{4(\beta \tri E)^{5/2}}{\pi^{3/2} l_0^2 L_c \Theta_c^4} \ee^{-\beta \tri E} 
\ . 
\eeq
Note that when $C_{\rm tot} \gg C_c$, we can ignore the monomeric concentration and $\bar{n}$ corresponds to the mean size of the polymers in the system  (\ref{eq:A5}). However, when $C_{\rm tot} \ll C_c$, $\bar{n} <1$ and it signifies that most particles are in the monomeric form \cite{carl_jphys97,liu_r17}.

\section{Details of simulation}
For the dimer, the longitudinal diffusion coefficient (DC)  is taken to be $\ln 2/4\pi$, the horizontal DC  $\ln 2/8\pi$, and the rotational DC $3/(8\pi \ln 2$) \cite{doi_b86}. Simulation is done in 2D and the dimension of the system is $[-4,4]\times [-1, 4]$, with periodic boundary condition in the $x$-direction and hard wall boundary condition in the $y$-direction. The dimer is initiated at the breakage point and then allowed to diffuse until reabsorbed by the $n$-mer, upon which it is immediately re-positioned to the breakage point. Results are from a simulation of $5\times 10^6$ time units. The model parameters are: $l_0=1$, $L_c=0.2$, $\Theta_c =0.5$ rad.

\section*{References}

\bibliographystyle{unsrt}

\providecommand{\newblock}{}

\end{document}